\documentclass[journal]{IEEEtran}

\ifCLASSINFOpdf
\else
   \usepackage[dvips]{graphicx}
\fi
\usepackage{url}

\usepackage{graphicx}
\usepackage{color, colortbl}
\usepackage{xcolor}
\definecolor{mygray}{gray}{0.9}
\usepackage{gensymb}
\usepackage{fixmath}
\usepackage{tablefootnote}
\usepackage{amssymb}
\usepackage{amsmath}
\usepackage{booktabs}
\usepackage{hyperref}

\begin{document}

\title{Neural Acoustic Context Field: Rendering Realistic Room Impulse Response With Neural Fields}

\author{Susan Liang, Chao Huang, Yapeng Tian, Anurag Kumar, and Chenliang Xu
\thanks{S.\ Liang, C.\ Huang, Y.\ Tian, and C.\ Xu are with the Department of Computer Science, University of Rochester, Rochester, NY, USA (e-mail: \{sliang22, chuang65\}@ur.rochester.edu, yapeng.tian@utdallas.edu, chenliang.xu@rochester.edu).}
\thanks{A.\ Kumar is with Meta Reality Labs Research, Redmond, WA, USA (e-mail: anuragkr90@meta.com).}
}

\markboth{Journal of \LaTeX\ Class Files, Vol. 14, No. 8, August 2015}
{Shell \MakeLowercase{\textit{et al.}}: Bare Demo of IEEEtran.cls for IEEE Journals}
\maketitle

\begin{abstract}
Room impulse response (RIR), which measures the sound propagation within an environment, is critical for synthesizing high-fidelity audio for a given environment. Some prior work has proposed representing RIR as a neural field function of the sound emitter and receiver positions. However, these methods do not sufficiently consider the acoustic properties of an audio scene, leading to unsatisfactory performance. This letter proposes a novel Neural Acoustic Context Field approach, called NACF, to parameterize an audio scene by leveraging multiple acoustic contexts, such as geometry, material property, and spatial information. Driven by the unique properties of RIR, i.e., temporal un-smoothness and monotonic energy attenuation, we design a temporal correlation module and multi-scale energy decay criterion. Experimental results show that NACF outperforms existing field-based methods by a notable margin. Please visit our project page for more qualitative results\footnote{\url{https://liangsusan-git.github.io/project/nacf/}}.
\end{abstract}


\IEEEpeerreviewmaketitle

\vspace{-8pt}
\section{Introduction}
\IEEEPARstart{H}{igh-fidelity} audio is essential in creating an immersive experience, as it enhances the audience's perception and engagement~\cite{raghuvanshi2018parametric,chaitanya2020directional}. For example in movies and video games, realistic sound effects are necessary for depicting believable virtual environments. In ``The Silence of the Lambs," for instance, 
the sound design reflects the terrifying prison environment, with metal doors clanging and footsteps echoing on concrete floors. Hence, generating audio with rich acoustic information is critical for the listener's perception of an environment.

Room impulse response (RIR), which characterizes the impact of room environment and emitter-receiver positions on sound propagation, is a valuable auditory function that aids in synthesizing audio with rich acoustic properties.
RIR consists of the direct propagation and early reflection parts, revealing the occlusion and distance, as well as the late reverberation component, conveying the scene size and structure.
By convolving an anechoic sound with the RIR signal, we can synthesize targeted audio that imitates the audio we would hear in this environment.
In essence, RIR offers rich acoustic cues that enable the receiver to discern the sound source position and approximate geometry of the surroundings.

The research on RIR can be traced back to the 1970s. Krokstad \textit{et al.}~\cite{krokstad1968calculating} and Vorl{\"a}nder~\cite{vorlander1989simulation} propose ray tracing algorithms to simulate sound propagation in rooms. Allen and Berkley~\cite{allen1979image} use a time-domain image expansion method for small-room acoustics simulation, and Borish~\cite{borish1984extension} extends this image model to arbitrary polyhedra with any number of sides. Recently, several wave-based methods~\cite{gumerov2009broadband,raghuvanshi2009efficient,hamilton2017fdtd} have been proposed to calculate RIR using the wave equation. However, these methods either require expensive computation or generate RIR with limited realism \cite{funkhouser2003survey,savioja2015overview}. 

Instead, this paper proposes an effective Neural Acoustic Context Field (NACF) approach for RIR generation by implicitly representing an indoor audio scene with neural fields. Specifically, NACF learns a mapping function from the positions of the sound emitter and receiver to the desired RIR to parameterize an audio scene, akin to traditional wave field coding methods~\cite{raghuvanshi2014parametric,chaitanya2020directional,raghuvanshi2018parametric}. The learned fields can then be queried with any emitter and receiver positions of interest for RIR generation. 
%
Our inspiration comes from the visual~\cite{nerf,martin2021nerf,li2022neural} and auditory~\cite{naf,inras} neural fields. 

However, applying neural fields na\"ively to audio scene representation results in unsatisfactory performance due to the lack of consideration of an audio scene's acoustic contexts or the RIR signal's unique properties, i.e., the energy attenuation and temporal un-smoothness characteristics. 
To address these limitations, (1) we introduce an acoustic context module to utilize different acoustic contexts, such as the geometry of an environment, material properties of the room surface, and the positions of the room boundary. 
(2) Considering the temporal un-smoothness characteristic of RIR at the direct propagation and early reflection stages, we further design a temporal correlation module to prevent smooth predictions. 
(3) Moreover, we propose a multi-scale energy decay criterion to supervise the RIR generation, which matches the monotonic energy attenuation trend of ground-truth RIR with the predicted counterpart to improve fidelity~\cite{fewshotrir}. 

\begin{figure*}[htbp]
    \centering
    \includegraphics[width=\textwidth]{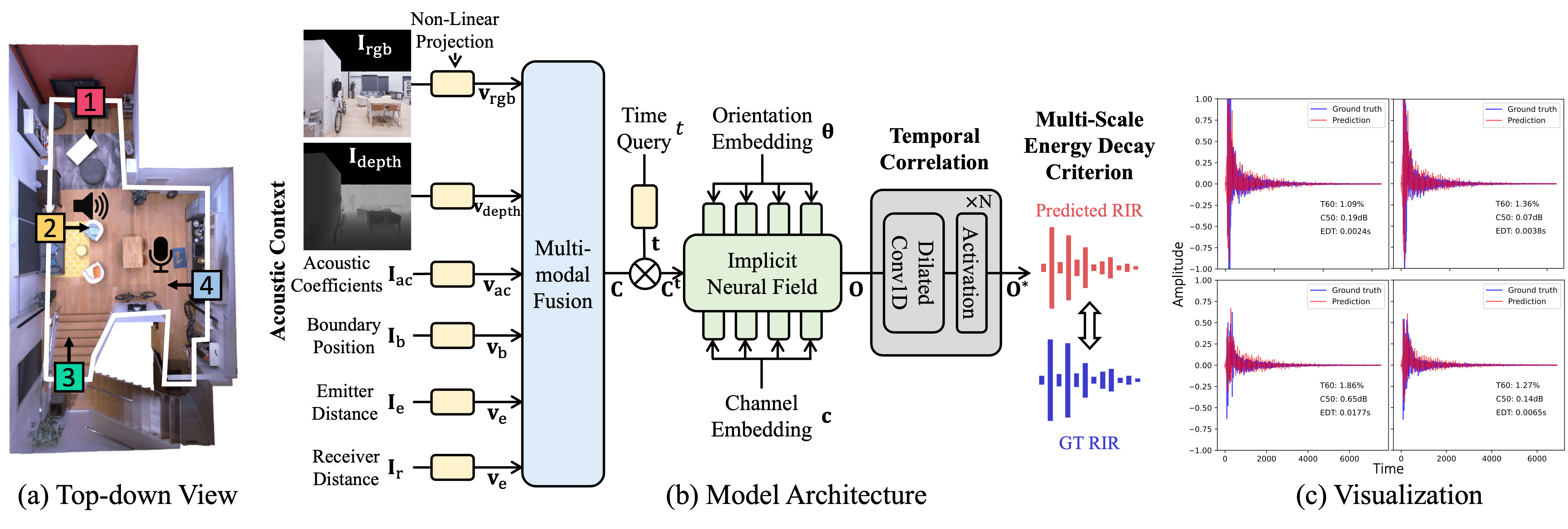}
    \vspace{-15pt}
    \caption{Method overview. The left (a) is the top-down view of an example indoor scene. We sample points evenly along the room boundary and extract various contextual information at each point, such as the RGB image, depth image, the acoustic coefficients of the surface, and several spatial information. The middle (b) is the architecture of our NACF model. First, we feed multiple acoustic contexts extracted along the room boundary (a) into the multi-modal fusion module. Then we integrate the fused contextual information with the time query as the spatial-temporal query, which is the input to the implicit neural field. After the neural field generates the RIR, we utilize a temporal correlation module to refine the RIR. Finally, we adopt the multi-scale energy decay criterion to supervise the model training. The right (c) is the visualization of predicted and ground-truth RIR together with generation errors.}
    \label{fig:model}
    \vspace{-15pt}
\end{figure*}


\section{Related Work}
\subsection{RIR Generation}
There are several directions in traditional RIR generation works, including ray tracing~\cite{krokstad1968calculating,vorlander1989simulation,cao2016interactive}, image source~\cite{martinez2010low,sprunck2022gridless}, wave equation~\cite{gumerov2009broadband,raghuvanshi2009efficient,chen2008analytical,bilbao2019local}, wave field coding~\cite{raghuvanshi2018parametric,chaitanya2020directional}, and sound field recording~\cite{ueno2017sound} methods. Recently, several methods that incorporate both geometric and material information~\cite{schissler2017acoustic,360audio,soundspaces,tang2022gwa} have appeared.
Unlike these simulation-based methods, learning-based methods leverage the powerful modeling ability of neural networks to analyze room acoustics and estimate RIR signals. Tang \textit{et al.}~\cite{tang2020scene} use deep neural networks to estimate the reverberation time and equalization of the room from recorded audio. ~\cite{ratnarajah2022fast,ratnarajah21_interspeech} use the generative adversarial network (GAN) to supervise realistic RIR generation. Estimating RIR from noisy audio~\cite{richard2022deep} or with few-shot training samples~\cite{majumder2022fewshot} has also been studied.

\vspace{-8pt}
\subsection{Implicit Neural Fields.}
In 2020, Mildenhall \textit{et al.}~\cite{nerf} proposed Neural Radiance Field (NeRF) for novel-view synthesis. Because of its powerful scene representation capability, NeRF and its variants soon gained popularity and are applied to different research domains, including vision~\cite{multivideonerf,lee2023dense}, audio~\cite{naf,inras}, and audio-visual~\cite{liang2023av}. Among them, NAF~\cite{naf} and INRAS~\cite{inras} are the most relevant methods to ours. NAF and INRAS leverage neural fields to model and render RIR. However, these methods do not sufficiently consider the acoustic information within an environment, leading to unsatisfactory performance.

\section{Method}
\subsection{Task Definition}
This letter targets rendering room impulse response with implicit neural fields. Given the 2D positions of the sound emitter $\mathbf{e} \in \mathbb{R}^2$ and receiver $\mathbf{r} \in \mathbb{R}^2$, the orientation of the sound receiver $\theta \in [0\degree, 360\degree)$, and the room impulse response signal $\mathbf{O} \in \mathbb{R}^{T \times 2}$, where $T$ is the length of the signal, and $2$ is the number of channels (we use two-channel binaural sound), our goal is learning a neural field $f: (\mathbf{e}, \mathbf{r}, \theta) \rightarrow \mathbf{O}$. Afterward, we can render RIR by querying the learned field with positions and orientations of interest, including queries in the training set and unobserved (novel) queries. 

\vspace{-8pt}
\subsection{Approach Overview}
As shown in Fig.~\ref{fig:model}, we enhance the RIR rendering capability of neural fields through three key components: acoustic context module, temporal correlation module, and multi-scale energy decay criterion. The acoustic context module provides a comprehensive understanding of the room acoustics to the neural fields, while the temporal correlation module prevents overly smooth predictions. Additionally, the loss criterion can reinforce the energy attenuation tendency of predicted RIR at different time-frequency resolutions.

\vspace{-8pt}
\subsection{Acoustic Context}
Sound propagation is mainly determined by (1) the geometry of an environment, (2) the material properties of the surface, and (3) the positions of the sound emitter and receiver. Therefore, we design an acoustic context module to encode all related acoustic information for RIR generation. 

Specifically, we sample $N$ points evenly along the room boundary for context extraction ($N=4$ in Fig.~\ref{fig:model}~(a)). We extract the indoor depth image $\mathbf{I}_{\mathrm{depth}} \in \mathbb{R}^{H\times W}$ and RGB image $\mathbf{I}_{\mathrm{rgb}} \in \mathbb{R}^{H\times W \times 3}$ from each point, where $H$ and $W$ are the height and width of the image, respectively. The depth image $\mathbf{I}_{\mathrm{depth}}$ depicts the geometry in the local region of each boundary point, while the RGB image $\mathbf{I}_{\mathrm{rgb}}$ contains the semantic information of different objects and can indicate the material property of the surrounding surface. We further extract the acoustic coefficients $\mathbf{I}_{\mathrm{ac}} \in \mathbb{R}_{+}^{P \times 3}$ of each boundary point that measures sound absorption, scattering, and transmission effects of $P$ main frequencies. To capture the spatial information within the room, we record the position of each point $\mathbf{I}_{\mathrm{b}} \in \mathbb{R}^{2}$, the distance between the emitter and each boundary point $\mathbf{I}_{\mathrm{e}} \in \mathbb{R}^{2}$, and the distance between the receiver and each boundary point $\mathbf{I}_{\mathrm{r}} \in \mathbb{R}^{2}$.  After capturing all acoustic contexts, we embed them into latent vectors of dimension $h$ ($\mathbf{v}_{\mathrm{depth}}, \mathbf{v}_{\mathrm{rgb}}, \mathbf{v}_{\mathrm{ac}}, \mathbf{v}_{\mathrm{b}}, \mathbf{v}_{\mathrm{e}}, \mathbf{v}_{\mathrm{r}} \in \mathbb{R}^h$). Consequently, we obtain six acoustic contexts from each boundary sample (Fig.~\ref{fig:model} (b)).

We then feed all contextual information into the multi-modal fusion module for acoustic context aggregation. In detail, we use the concatenation operator to fuse acoustic contexts from different modalities and various boundary points as the holistic knowledge of room acoustics $\mathbf{C} \in \mathbb{R}^{N \times 6 \times h}$:
\begin{equation}
    \begin{aligned}
    \mathbf{C} &= [\mathbf{C}^1,\mathbf{C}^2,\cdots,\mathbf{C}^N]\enspace, \\
    \mathbf{C}^i = [\mathbf{v}_{\mathrm{depth}}^i &, \mathbf{v}_{\mathrm{rgb}}^i , \mathbf{v}_{\mathrm{ac}}^i , \mathbf{v}_{\mathrm{b}}^i , \mathbf{v}_{\mathrm{e}}^i , \mathbf{v}_{\mathrm{r}}^i],\ 1 \leq i \leq N \enspace, \\
    \end{aligned}
\end{equation}
where $\mathbf{C}^i \in \mathbb{R}^{6 \times h}$ is the contextual information extracted from the boundary point $i$, $N$ is the number of boundary points, and $6$ is the number of context categories.

\vspace{-8pt}
\subsection{Implicit Neural Field}
Once the acoustic context $\mathbf{C}$ has been estimated, we proceed to integrate it with the time query $t \in [1, T]$. Similar to NeRF~\cite{nerf}, we employ positional encoding $\gamma$ to project the single-value time query $t$ to a high-dimension space $\gamma(t) \in \mathbb{R}^{2L}$, where $L$ is the number of frequencies. Subsequently, we non-linearly embed $\gamma(t)$ into a time vector $\mathbf{t} \in \mathbb{R}^h$. Finally we calculate the dot product between the acoustic context $\mathbf{C}$ and the time vector $\mathbf{t}$, resulting in the modified time-aware acoustic context $\mathbf{C}_t \in \mathbb{R}^{N \times 6}$: 
\begin{equation}
    \mathbf{C}^{(i,j)}_{t} = \mathbf{C}^{(i,j)} \cdot \mathbf{t},\ 1 \leq i \leq N,\ 1 \leq j \leq 6 \enspace.
\end{equation}
By incorporating the spatial context ($\mathbf{C}$) and the temporal information ($\mathbf{t}$), the modified acoustic context $\mathbf{C}_t$ can effectively serve as a spatial-temporal query for implicit neural fields.

Next, we feed $\mathbf{C}_t$ into our implicit neural field, which is instantiated as an MLP network $\Theta$, to generate the RIR signal:
\begin{equation}
    f_\Theta: (\mathbf{C}_t, \theta, c) \rightarrow \mathbf{O}_{t,c} \enspace,
\end{equation}
where $\theta$ is the listener head orientation, $c$ is the channel index of audio signals ($c$ can be left or right for binaural audios), and $\mathbf{O}_{t,c}$ is the RIR of time $t$ and channel $c$. To obtain the complete RIR signal $\mathbf{O}$, we query the learned neural field with all spatial-temporal queries $\mathbf{C}_{1:T}$ and all channels $c$.

To ease the optimization process of implicit neural fields and enhance the modeling ability, we use learnable orientation embeddings $\mathbold{\theta} \in \mathbb{R}^h$ and channel embeddings $\mathbf{c} \in \mathbb{R}^h$ to replace the orientation $\theta$ and the channel $c$, respectively. We add $\mathbold{\theta}$ and $\mathbf{c}$ to the input of all layers of the neural field as the orientation and channel conditions.

\vspace{-8pt}
\subsection{Temporal Correlation}
Once the complete RIR signals $\mathbf{O}$ have been generated, we adopt a temporal correlation module to filter and refine the RIR signals. The design is motivated by the continuity of neural fields: similar input queries yield similar output predictions. This property results in smooth RIR signals estimated by implicit neural fields. However, RIR signals typically are temporally un-smooth at the direct propagation and early reflection stages. Therefore, we introduce a temporal correlation module to prevent smooth predictions. 

As shown in Fig.~\ref{fig:model}, we append the neural field with several layers of dilated one-dimension convolution neural network $g_{\Phi}$, each of which is followed by an activation layer. We denote the refined RIR signals as $\mathbf{O}^{*}$ with $g_{\Phi}: \mathbf{O} \rightarrow \mathbf{O}^{*}$.

\vspace{-8pt}
\subsection{Multi-scale Energy Decay Criterion}
Finally, we optimize NACF to predict realistic RIR with two supervision signals. Given a two-channel RIR signal $\mathbf{O}$ (or $\mathbf{O}^*$ after the temporal correlation), we use Short Time Fourier Transformation (STFT) to convert it from the time domain to the time-frequency domain and calculate its magnitude $\mathbf{M} \in \mathbb{R_{+}}^{F \times D \times 2}$, where $F$ is the number of frequency bins, $D$ is the number of time windows, and $2$ denotes the two channels. We calculate the magnitude of ground-truth RIR $\mathbf{M}_{g}$ and that of predicted RIR $\mathbf{M}_{p}$, and measure their L1 distance as the first training objective: 
\begin{equation}
    \mathcal{L}_{\mathrm{mag}} = ||\mathbf{M}_{g} - \mathbf{M}_{p}||_1 \enspace.
\end{equation}

To reinforce the energy attenuation tendency of predicted RIR, we follow Majumder \textit{et al.}~\cite{fewshotrir} using energy decay matching loss as the second training objective. Given a magnitude $\mathbf{M}$, we first compute its energy in each time window by calculating the magnitude's square and aggregating it along the frequency dimension:
\begin{equation}
    \mathbf{M}'^{(d)} = \sum_{f=1}^{F} \left(\mathbf{M}^{(f,d)} \right)^2,\ 1 \leq d \leq D \enspace,
\end{equation}
where $\mathbf{M}' \in \mathbb{R}_{+}^{D \times 2}$ represents the energy in each time window.

We then sum $\mathbf{M}'$ along the time dimension, aggregating the energy from the current step $d$ until the end $D$ for each time step $d \in [1, D]$ to capture the overall energy decay trend:
\begin{equation}
    \mathbf{M}''^{(d)} = \sum_{i=d}^{D} \mathbf{M}'^{(i)},\ 1 \leq d \leq D \enspace,
\end{equation}
where $\mathbf{M}'' \in \mathbf{R}_{+}^{D \times 2}$. The resulting $\mathbf{M}''$ has the same shape as $\mathbf{M}'$ since we calculate the energy sum for all time steps $d$. Finally, we measure the L1 distance between the ground-truth energy trend and the predicted one in the log space:
\begin{equation}
    \mathcal{L}_{\mathrm{dcy}} = ||\log_{10}\mathbf{M}''_g - \log_{10}\mathbf{M}''_p||_1 \enspace.
\end{equation}

\begin{figure*}[htbp]
    \centering
    \includegraphics[width=\textwidth]{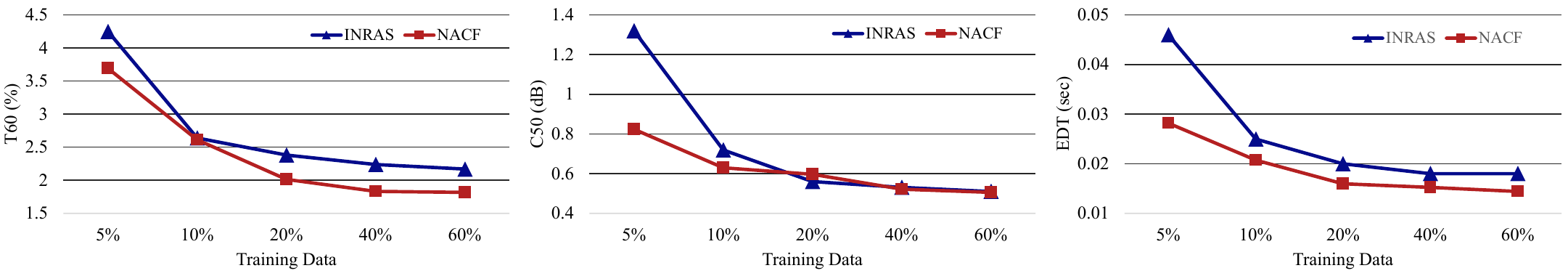}
    \caption{Few-shot learning. We report the results of NACF and INRAS trained on $5\%$ to $60\%$ data. Compared with INRAS, our model can achieve more robust performance when the training data is insufficient. Note that the few-shot results are reported in one representative scene instead of all six scenes presented in Table~\ref{tab:sota} because INRAS only reports its performance in this single scene (``frl\_apartment\_4").}
    \label{fig:fewshot}
    \vspace{-15pt}
\end{figure*}

Because both $\mathcal{L}_{\mathrm{mag}}$ and $\mathcal{L}_{\mathrm{dcy}}$ are computed in the time-frequency domain, their effectiveness relies on the chosen window sizes and frequency bins for STFT. Therefore, we use a set of window sizes $\{W_i\}$
and frequency bins $\{F_i\}$ to assess the prediction qualities at different scales. If we view $\mathcal{L}_{\mathrm{mag}}$ and $\mathcal{L}_{\mathrm{dcy}}$ as functions of window sizes and frequency bins, the overall loss can be expressed as:
\begin{equation}
    \mathcal{L} = \sum_{i=1}^{S} ( \mathcal{L}_{\mathrm{mag}}(W_i,F_i) + \lambda \cdot \mathcal{L}_{\mathrm{dcy}}(W_i,F_i) ) \enspace,
\end{equation}
where $\lambda$ is a hyper-parameter that controls the weight of $\mathcal{L}_{\mathrm{dcy}}$, and $S$ is the set size ($|\{W_i\}|$ and $|\{D_i\}|$).

\section{Experiments}
\subsection{Experimental Settings}
\noindent\textbf{Dataset.} We evaluate our Neural Acoustic Context Field using the SoundSpaces dataset~\cite{soundspaces}. The SoundSpaces dataset consists of extensive high-quality RIR computed in simulation environments. For each sound emitter and receiver pair, SoundSpaces provides binaural (two-channel) RIR for four discrete head orientations (0, 90, 180, and 270 degrees). To ensure a fair comparison, we adopt the same six representative scenes for training and evaluation, as in previous works~\cite{naf,inras}: two single rooms with rectangular walls, two single rooms with non-rectangular walls, and two multi-room layouts. We maintain the same training/test split as NAF~\cite{naf} with 90\% data for training and 10\% data for testing. For further dataset details, please refer to NAF. We utilize the Habitat-Sim simulator \cite{szot2021habitat,habitat19iccv} to extract various acoustic contexts, such as RGB images, depth images, and acoustic coefficients.

\noindent\textbf{Metrics.} Following INRAS~\cite{inras}, we select three metrics, namely T60, C50, and EDT, to assess the RIR generation quality of our model. T60 measures the time it takes for the energy to decay by $60$ dB. C50 captures the energy ratio between the first $50$ms of RIR and the remaining portion. EDT is similar to T60 but focuses more on the early reflection of RIR. Please refer to IRNAS for details of these metrics.

\noindent\textbf{Architectures.} We implement our implicit neural field as a 4-layer 256-width MLP with one skip connection layer. For the temporal correlation module, we stack three 1-dimension convolution layers of dilation size $2$ and gradually increase the kernel size from $3$ to $7$ to enlarge the receptive field. For all non-linear projections, we use 2-layer 256-width MLPs. We set $L=10$ for positional encoding $\gamma(t)$. The orientation $\mathbold{\theta}$ and channel embeddings $\mathbf{c}$ are randomly initialized as $256$-dimension vectors.

\noindent\textbf{Training Details.} All models are trained using Adam~\cite{adam} optimizer for $100$ epochs with a batch size of $32$ and a learning rate of $5e{-}4$. To facilitate training, we adopt curriculum learning~\cite{bengio2009curriculum} by initially training the model without the temporal correlation module (referred to as \textbf{NACF}) and then separately training the temporal correlation module with the main model frozen (\textbf{NACF w/ T}). We set $\{W_i\} = \{240, 600, 1200\}$, $\{F_i\} = \{512, 1024, 2048\}$, and $\lambda = 0.01$.

\vspace{-8pt}
\subsection{Evaluation}
\noindent\textbf{Results.} We compare our model with existing similar works, including the state-of-the-art method INRAS~\cite{inras}. In line with INRAS, we include the results of traditional audio encoding methods, such as Advanced Audio Coding (AAC) \cite{aac} and Xiph Opus \cite{opus}, since our model can be interpreted as an audio encoding approach. To evaluate the quality of our generated RIR, we employ the T60, C50, and EDT metrics, where a lower score indicates better RIR quality. As depicted in Table \ref{tab:sota}, \textbf{NACF} outperforms all other approaches by significant margins across all metrics. Compared to INRAS, \textbf{NACF} reduces the T60 error by $0.78$, the C50 error by $0.1$ dB, and the EDT error by $0.0047$ sec. Additionally, incorporating the temporal correlation module further boosts performance: \textbf{NACF w/ T} achieves over 31\% relative improvement on the T60 metric, 18\% relative improvement on C50, and 26\% on EDT compared to INRAS.
\begin{table}[htbp]
\vspace{-5pt}
\centering
\caption{Comparison with the SOTA. We report the performance on the SoundSpaces dataset using T60, C50, and EDT metrics. A lower score indicates higher RIR generation quality.}
\begin{tabular}{l|ccccc}
\toprule
Methods      & T60 (\%) $\downarrow$  & C50 (dB) $\downarrow$ & EDT (sec) $\downarrow$  \\
\hline
Opus-nearest & 10.10 & 3.58 & 0.115 \\
Opus-linear  & 8.64  & 3.13 & 0.097 \\
AAC-nearest  & 9.35  & 1.67 & 0.059 \\
AAC-linear   & 7.88  & 1.68 & 0.057 \\
\hline
NAF \cite{naf}         & 3.18  & 1.06 & 0.031 \\
INRAS \cite{inras}       & 3.14  & 0.60 & 0.019 \\
\textbf{NACF (Ours)}         & \textbf{2.36}  & \textbf{0.50} & \textbf{0.014$\mathbf{_3}$} \\
\rowcolor{mygray}
\textbf{NACF w/ T} & \textbf{2.17} & \textbf{0.49} & \textbf{0.013$\mathbf{_9}$} \\
\bottomrule
\end{tabular}
\label{tab:sota}
\vspace{-8pt}
\end{table}

\noindent\textbf{Ablation Studies.} We break down our model to analyze the contributions of different components toward the final performance. We report the ablation results in one representative scene (``apartment\_2") in Table \ref{tab:ablation}: the inclusion of the temporal correlation module can boost the performance, while the exclusion of either acoustic context or multi-scale energy decay criterion degrades the generation quality.
\begin{table}[htbp]
\vspace{-5pt}
\centering
\caption{Ablation studies. We break down our model to analyze the influence of different components on the generation quality. NACF w/ T means NACF with the temporal correlation module, NACF w/o C is without the acoustic context module, and NACF w/o M is without multi-scale energy decay criterion.}
\begin{tabular}{l|ccc}
\toprule
Methods      & T60 (\%) $\downarrow$  & C50 (dB) $\downarrow$ & EDT (sec) $\downarrow$  \\
\hline
NACF & 2.33 & 0.55 & 0.0162\\ 
\hline
\rowcolor{mygray}
NACF w/ T & \textbf{2.25} & \textbf{0.54} & \textbf{0.0160}\\ 
NACF w/o C & 2.96 & 0.75 & 0.0183\\ 
NACF w/o M & 5.15 & 0.98 & 0.0255\\ 
\bottomrule
\end{tabular}
\label{tab:ablation}
\end{table}

\noindent\textbf{Visualization.} For the intuitive perception of the predicted RIR, we visualize the waveform of RIR in Fig.~\ref{fig:model} (c).

\noindent\textbf{Few-shot Learning.} We also examine the reliance of our model on the data scale by conducting a few-shot learning experiment in the same scene as INRAS. We feed our model with only $5\%$ of the training data and gradually increase the portion to $60\%$ and test NACF's performance. Fig.~\ref{fig:fewshot} demonstrates that our model can achieve more robust performance than INRAS even when the training data is insufficient.

\section{Conclusions}
This letter proposes a novel method of rendering room impulse response called NACF. NACF utilizes the contextual information of the room acoustics to generate realistic RIR signals robustly. With the aid of the acoustic neural field, temporal correlation module, and multi-scale energy decay criterion, NACF outperforms previous work with a clear margin and sets the new state-of-the-art performance on the SoundSpaces dataset.

\newpage
\small{
\bibliographystyle{abbrv}
\bibliography{mybib}
}
\end{document}